# Structural Stability and Optoelectronic Properties of Tetragonal MAPbI$_3$ under Strain

Lei Guo[1], Gao Xu[2], Gang Tang[1], Daining Fang[2,3], Jiawang Hong[1*]

[1]School of Aerospace Engineering, Beijing Institute of Technology, Beijing, 100081, PR China.

[2]State Key Laboratory for Turbulence and Complex Systems, College of Engineering, Peking University, Beijing, 100871, PR China.

[3]Institute of Advanced Structure Technology, Beijing Institute of Technology, Beijing 100081, PR China

E-mail: hongjw@bit.edu.cn

**ABSTARCT**

In recent years, organic-inorganic hybrid perovskites have attracted wide attention due to their excellent optoelectronic properties in the application of optoelectronic devices. In the manufacturing process of perovskite solar cells, perovskite films inevitably have residual stress caused by non-stoichiometry components and the external load. However, their effects on the structural stability and photovoltaic performance of perovskite solar cells are still not clear. In this work, we investigated the effects of external strain on the structural stability and optoelectronic properties of tetragonal $MAPbI_3$ by using the first-principles calculations. We found that the migration barrier of $I^-$ ion increases in the presence of compressive strain and decreases with tensile strain, indicating that the compressive strain can enhance the structural stability of halide perovskites. In addition, the light absorption and electronic properties of $MAPbI_3$ under compressive strain are also improved. The variations of the band gap under triaxial and biaxial strains are consistent within a certain range of strain, resulting from the fact that the band edge positions are mainly influenced by the Pb-I bond in the equatorial plane. Our results provide useful guidance for realizing the commercial applications of $MAPbI_3$-based perovskite solar cells.

## 1. Introduction

In the past several years, the power conversion efficiency (PCE) of organic-inorganic hybrid perovskites (OIHPs) has increased from 3.81% to 25.2%,[1-5] which attracts tremendous attention on the understanding of superior photovoltaic performances and various potential optoelectronic applications. It is especially noteworthy that OHIPs possess good flexibility and are very suitable to serve as energy storage devices for flexible screens, software robots and wearable devices.[6-8] In practical applications, OIHPs are inevitably affected by external loads such as tension, bending and torsion, *etc*, which makes the crystal structure easily decomposed. These factors result in a short service lifetime and limit their commercialization.[9]

Various methods have been used to improve the stability of OIHPs' crystal structure, such as doping, substitution of heterogeneous elements and structural packaging, *etc*.[10-14] Han *et al.* used full-printing method to prepare a mesoporous structure perovskite solar cells with stability greater than 1000 h, however, it is still difficult to meet the requirements of commercial applications.[15] Huang *et al.* experimentally found that when $CH_3NH_3PbI_3$ (MA = $CH_3NH_3$) is grown on the substrate, since the thermal expansion coefficient of $MAPbI_3$ is larger than the thermal expansion coefficient of the substrate, residual stress is left in $MAPbI_3$ after annealing, resulting in a decrease in the lifetime of $MAPbI_3$ devices, which may be contributed to the decrease of the ion migration barrier with the presence of tensile strain.[16] Rolston *et al.* applied compressive strain on the crystal structure of $MAPbI_3$ and found that the device lifetime increased under illumination.[17] In 2019, Chen *et al.* revealed the evolution mechanism of the residual strain of the mixed halide perovskite $(FAPbI_3)_{0.85}(MAPbBr_3)_{0.15}$ (FA = $CH(NH_2)_2$) in the vertical direction, and developed a simple technique to adjust the distribution of the strain and optimize the PCE.[18] This opens up a new path for us to tune the crystal stability of OHIPs. However, the influence of external strain on the crystal structure stability and the ion migration of OHIPs under external strain still lack in-depth systematic research, and the intrinsic mechanism also needs to be further explored. Moreover, the optoelectronic properties of OHIPs under strain also urgently need to be further theoretically studied, so as to better control the structure and optoelectronic properties of OHIPs by using simple mechanical means.

In this work, taking the traditional organic-inorganic hybrid perovskite MAPbI$_3$ (tetragonal phase, *I4/mcm*) as an example, the effects of strain on the structural stability and optoelectronic properties of MAPbI$_3$ were investigated by using the first-principles calculations. We found that the migration barrier of I$^-$ ion increases in the presence of compressive strain and decreases with tensile strain, indicating that the compressive strain can enhance the crystal structural stability of this material. Our results give a reasonable explanation of the increased lifetime of perovskite solar cells observed in the experiment. And it is also beneficial for increasing the light absorption and PCE by applying compressive strain on MAPbI$_3$. In addition, the variations of the band gap under triaxial and biaxial strains are consistent within a certain range of strain, resulting from the fact that the band edge positions are mainly influenced by the Pb-I bond in the equatorial plane.

## 2. Computational details

Our calculations are performed by using the Vienna *Ab-initio* Simulation Package (VASP) code in the framework of density function theory (DFT).[19, 20] The electron-ion interaction is described by the projector augmented wave (PAW) method.[21] The generalized gradient approximation (GGA) of Perdew-Burke-Ernzerhof (PBE) is used to describe the exchange-correlation function.[22] The vdW-DF3 method was used to consider the Van der Waals (vdW) interactions, which plays an important role in the hybrid perovskites materials with weak interaction along the stacking direction.[23, 24] The electronic orbitals $5d^{10}6s^26p^2$, $5s^25p^5$, $2s^22p^2$, $2s^22p^3$, and $1s^1$ are considered in valence for Pb, I, C, N, and H atoms, respectively. The plane-wave cut-off energy was set to 650 eV. The 6×6×6 Monkhorst-Pack *k*-point mesh[25] was employed for sampling the Brillouin zone. Both *k*-point mesh and the cut-off energy were tested carefully for the convergence. The lattice parameters and atomic positions were fully relaxed until the energy difference is less than 10$^{-5}$ eV and the force on each atom is smaller than $10^{-2}$ eV/ Å. Under triaxial strain (the strain is the same), the lattice constants of *a*, *b* and *c* axes are fixed and then only the atomic positions are relaxed. Under biaxial strain (strain along *a* and *b* directions and strain is the same), we fixed the lattice constants of *a* and *b* axes after applying strain, and the lattice constant *c* and atomic positions are fully relaxed.

The activation energy ($\triangle E_a$) is calculated by using the Nudged Elastic Band (NEB) method.[26, 27] The 2×2×2 supercell of MAPbI$_3$ (including the strained crystal structures) was used, and the start and end points were constructed by removing the I atom from the perfect crystal structure. A discretized path with 7 images is used during the NEB calculations, including start and end points. Only the 1×1×1 *k* points were employed to perform the geometry optimization of supercell. $\triangle E_a$ was obtained as the difference between the energies of the maximum and start points of the calculated NEB images.

The optical properties of MAPbI$_3$ are described by the complex dielectric function, *i.e.* $\varepsilon(\omega) = \varepsilon_1(\omega) + i\varepsilon_2(\omega)$. Based on the obtained dielectric function of MAPbI$_3$, the absorption coefficient $\alpha(\omega)$ can be given by the following equation:

$$\alpha(\omega) = \sqrt{2}\omega\sqrt{\sqrt{\varepsilon_1(\omega)^2 + \varepsilon_2(\omega)^2} - \varepsilon_1(\omega)} / c$$

where $\varepsilon_1$ and $\varepsilon_2$ are the real and imaginary part of the dielectric function, respectively.[28]

## 3. Geometry structure

According to the experimental results of Poglitsch and Weber, organic-inorganic hybrid perovskite MAPbI$_3$ adopts three kinds of crystal structures such as orthorhombic phase, tetragonal phase and cubic phase.[29] In this work, the tetragonal structure (see Fig. 1) at room temperature of MAPbI$_3$ was chosen to investigate the structural stability and optoelectronic properties under strain. From Fig. 1, the center of octahedron is occupied by Pb atom, the I atoms share the corner positions of octahedron, and the organic cation $MA^+$ lies in a cage surrounded by [$PbI_6$] octahedron. We fully relaxed the atomic coordinates and lattice parameters to obtain the ground state structure, the relaxed lattice constants are shown in Table I. The lattice constants of *a*-axis and *b*-axis show a slight difference with 0.03 due to the influence of the orientation of organic molecule MA, and the calculated results are in good agreement with the experimental values (the error is less than 2%).[29]

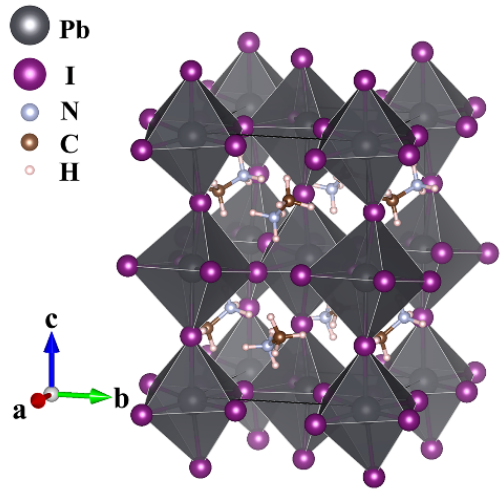

Figure 1. The tetragonal phase structure of $MAPbI_3$.

Table I. Calculated and experimental[29] lattice parameters of tetragonal phase of $MAPbI_3$.

| $MAPbI_3$ | *a* () | *b* () | *c* () |
|---|---|---|---|
| This work | 8.69 | 8.72 | 12.83 |
| Experiment | 8.86 | 8.86 | 12.66 |
| Error (%) | 1.92 | 1.58 | 1.34 |

The lattice constants of $MAPbI_3$ under strain are shown in Fig. 2. Under the triaxial strain, the lattice constants (*a*, *b* and *c*) increase with the increase of strain. While under the biaxial strain, the lattice constants (*a* and *b*) increase gradually, while the lattice constant (*c*) decreases gradually from -4% to 4% biaxial strain due to the Poisson's ratio effect.

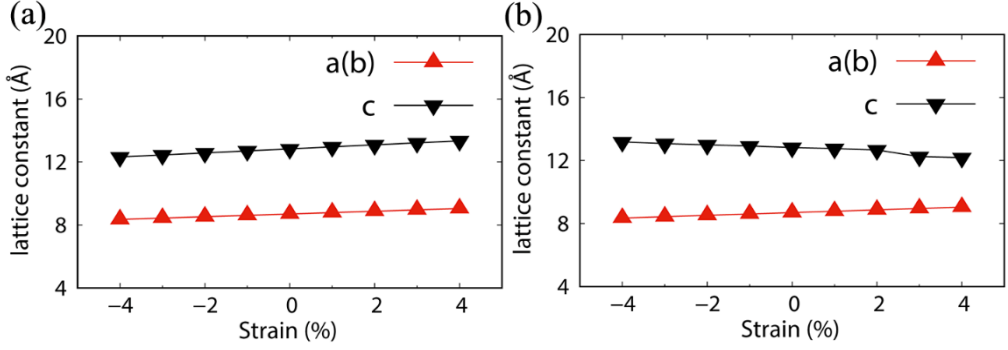

Figure 2. The lattice constants of MAPbI$_3$ under triaxial strain (a) and biaxial strain (b).

In order to study the changes of bond lengths and bond angles in the strained MAPbI$_3$ structure, we chose the corresponding bond lengths and bond angles in the equatorial plane (e: equatorial plane) and apical plane (a: apical) as representatives. $l_e$ and $l_a$ represent the average lengths of all Pb-I bonds in the equatorial plane and all Pb-I bonds in the apical plane respectively. While $α_e$ and $α_a$ represent the average angles of Pb-I-Pb angles in the equatorial plane and apical plane, respectively. The bond angle of Pb-I-Pb is used to measure the tilt degree of [PbI$_6$] octahedron.

The changes of average Pb-I bond lengths ( $l_e$ and $l_a$ ) and Pb-I-Pb bond angles ($α_e$ and $α_a$ ) under triaxial strains are shown in Fig. 3. Under triaxial compression strain, the Pb-I bond lengths ($l_e$ and $l_a$) in the crystal structure of MAPbI$_3$ becomes shorter (Fig. 3a) and the bond angle of Pb-I-Pb decreases (Fig. 3b), which means the [PbI$_6$] octahedron inclines more under compression. While under triaxial tension strain, it is opposite. Under 4% triaxial tension strain, however, the increase of Pb-I bond $l_e$ is small, and bond angle $α_a$ decreases (the sudden increase of $l_a$ leads to the structural damage of [PbI$_6$] octahedron, as shown in Fig. 3a). The variation of the bond angle of Pb-I-Pb under triaxial strain shows that the compressive strain increases the tilt of [PbI$_6$] octahedron, while the tensile strain decreases the tilt of [PbI$_6$] octahedron (with 4% strain, the tilt of [PbI$_6$] octahedron increases in apical plane direction due to the sudden increase of $l_a$ which may induced by the displacement of I ion, compared with 2% strain).

The changes of average Pb-I bond lengths ($l_e$ and $l_a$) and Pb-I-Pb bond angles ($\alpha_e$ and $\alpha_a$) under biaxial strains are shown in Fig. 4. Under biaxial compressive strain, $l_e$ becomes shorter but $l_a$ becomes longer (Fig. 4a). $\alpha_e$ becomes smaller while $\alpha_a$ becomes larger (Fig. 4b). Under biaxial tensile strain, it is opposite. However, above 2% biaxial tension strain, $l_a$ becomes longer, which makes $\alpha_a$ decrease sharply. This is because the distances of Pb atoms in apical plane direction become more closer which pushes the I atoms between Pb atoms in apical plane far away, increasing the Pb-I bond length $l_a$ and triggered a sharp decrease in $\alpha_a$. The variation of the bond angle of Pb-I-Pb under biaxial strain indicates that the compressive strain increases the tilt of [PbI$_6$] octahedron in the equatorial plane while decreases the tilt in the apical plane. It is opposite under tensile strain.

The analysis of the structure under triaxial and biaxial strains clearly shows that the changes of the bond angles of Pb-I and Pb-I-Pb in MAPbI$_3$ are consistent in the equatorial plane, but opposite in the apical plane.

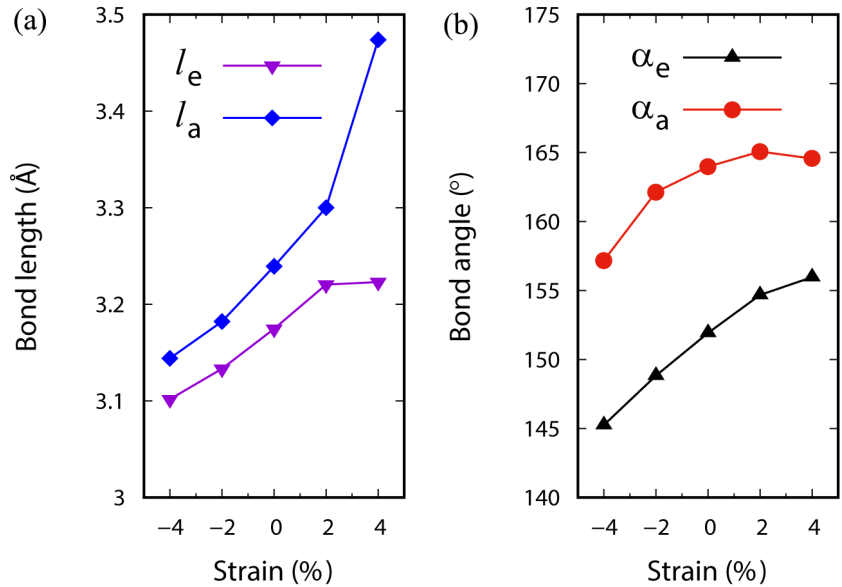

Figure 3. (a) the average Pb-I bond lengths $l_e$ and $l_a$ and (b) the average Pb-I-Pb bond angles $\alpha_e$ and $\alpha_a$ of MAPbI$_3$ under triaxial strains

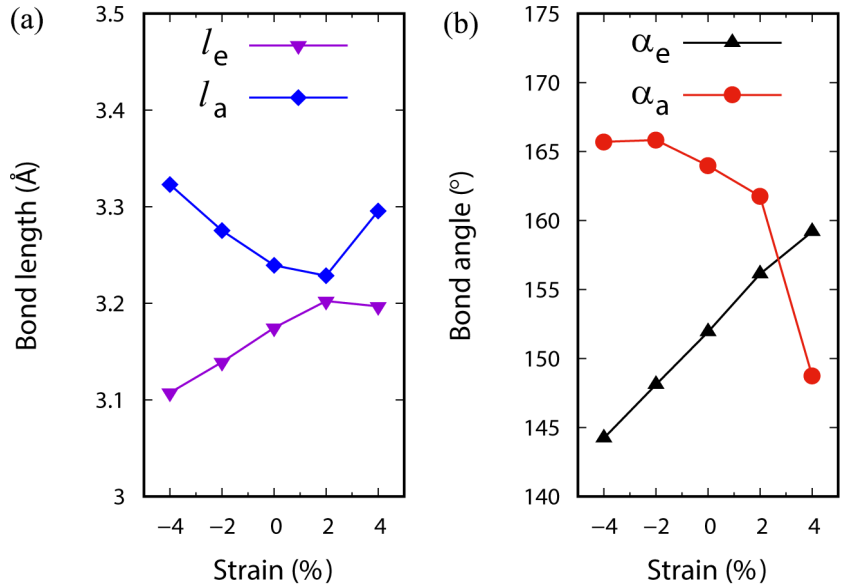

Figure 4. (a) the average Pb-I bond lengths $l_e$ and $l_a$ and (b) the average Pb-I-Pb bond angles $\alpha_e$ and $\alpha_a$ of MAPbI$_3$ under biaxial strains

## 4. Iodine Ion Migration under strain

Dauskardt *et al.* observed that compressive strain can prolong the lifetime of MAPbI$_3$-based perovskite solar cells, while tensile strain can reduce their lifetime.[30] Huang et al. observed that the residual stress in MAPbI$_3$ could lead to the decrease of the barrier of ion transport.[16] Here, we theoretically study the effect of the strain on the barrier of ion migration in organic-inorganic hybrid perovskite to understand the intrinsic mechanism.

Tateyama et al. calculated the migration barriers of MA molecule and I ion in MAPbI$_3$ crystal structure along different paths in the absence of strain by the first-principles calculations.[31] Compared with MA molecule, the migration barrier of I ion in equatorial plane is the lowest, indicating that it is easier to migrate. Therefore, we chose the I ion in the equatorial plane to study the change of migration barrier under strain. The migration path **P** is shown in Fig. 5.

We constructed a $2\times2\times2$ MAPbI$_3$ supercell (384 atoms) and calculated the migration barrier of I ions by using the NEB method. Firstly, we removed an I atom to construct the initial and final states, and inserted five transition states between the initial and final states. Then, we synergistically relaxed five transition states to obtain the corresponding energy of seven structures including the initial and final states. The migration barrier is the difference between the energy of the highest state and the initial state.

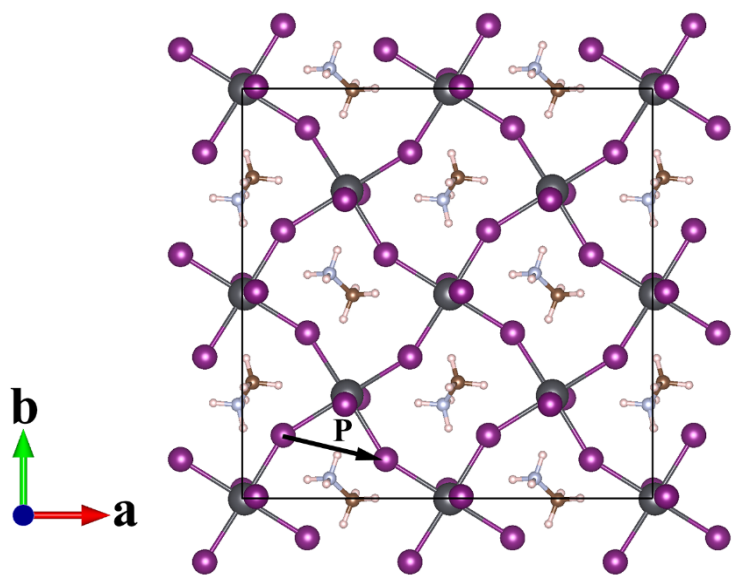

Figure 5. The migration path **P** of I ion in the equatorial plane

We calculated the migration barriers of I ion in the equatorial plane from -4% to 4% strain, as shown in Fig. 6 and Fig. 7. From -4% to 4% triaxial strain (Fig. 6), the migration barrier of I ion decreases gradually. Under biaxial strain (Fig. 7), the I ion migration barrier does not show obvious change at 2% tensile strain, and when the strain continues to increase, the barrier will decrease a lot. When compressive strain is applied, the migration barrier of I ions increases gradually.

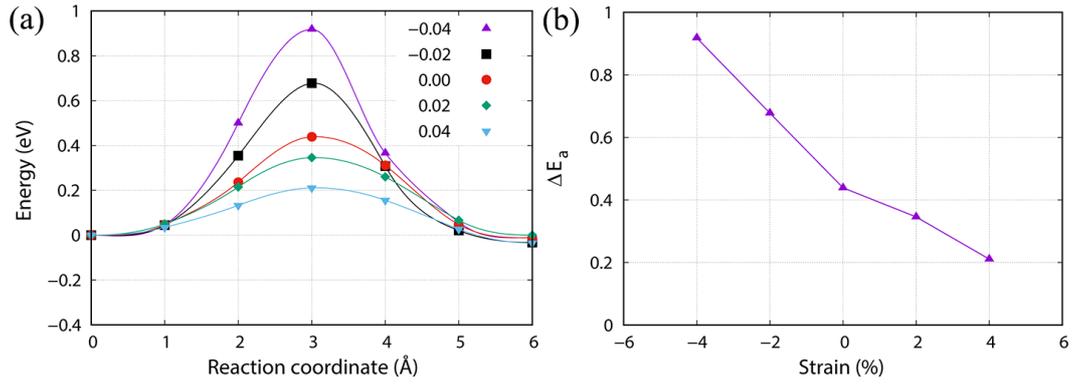

Figure 6. (a) Energy corresponding to different transition States in I ion migration under triaxial strain. (b) Migration barrier of I ion under triaxial strain

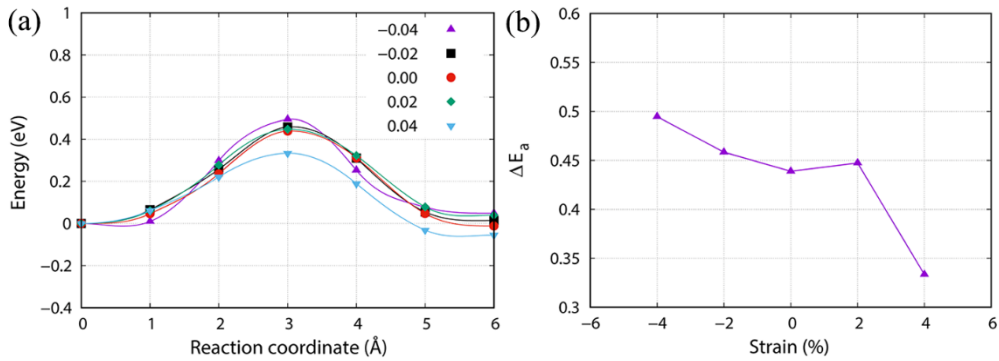

Figure 7. (a) Energy corresponding to different transition States in I ion migration under biaxial strain. (b) Migration barrier of I ion under biaxial strain

We can see that both triaxial and biaxial tensile strain reduce the migration barrier of I ions, while compressive strain increases the migration barrier from Fig. 6 and Fig. 7. This indicates that in MAPbI$_3$, stretching makes it easier for I ions to migrate in the equatorial plane, while compression hinders its migration in the equatorial plane. According to previous studies [32-37], the structure of MAPbI$_3$ is easy to decompose into two products such as MAI and PbI$_2$ under external environment (such as humidity, temperature, *etc*.), which results in aging phenomenon. Therefore, we believe that I ion migrates more easily under tension strain and will combine with organic molecules MA and Pb ions to form MAI and PbI$_2$, which damages the structure of MAPbI$_3$ and reduces the lifetime of MAPbI$_3$ film, whereas compressive strain has the opposite effect and improves the stability of MAPbI$_3$ film.

The reason why strain can change the migration barrier of I ion in MAPbI$_3$ is related to the structural changes. When strain is applied, the length of Pb-I bond and the angle of Pb-I-Pb bond change in MAPbI$_3$, and the rotation of [PbI$_6$] octahedron also occurs. For triaxial strain, compressive strain shortens the length of Pb-I bond and distorts the octahedron (as shown in Fig. 3), which enhances the interaction between Pb and I atom, making it difficult for I ion to escape from the potential field of Pb atom, increasing the migration barrier of I ion in the plane and hindering its migration in the plane. Tensile strain makes the length of Pb-I bond longer, and the interaction between them becomes weaker, thus I ions migrate more easily.

From the above discussions, we find that the migration barrier of I ion in MAPbI$_3$ can be tuned by simple strain method, and the structural stability can be improved in compressive strain. Next, we will investigate how strain affects the optoelectronic properties of MAPbI$_3$ which is essential for the photovoltaic applications.

## 5. Optoelectronic properties under strain

It is found that the structural stability of absorbers can be improved to prolong the device lifetime of perovskite solar cells in MAPbI$_3$ under compressive strain. [16-18] As photovoltaic materials, the light absorption capacity of MAPbI$_3$ is one of the important indicators to measure its performance. Therefore, it is of great significance to study the changes of light absorption of MAPbI$_3$ in visible region under strain. Katrasiak *et al.* found that with the increase of hydrostatic pressure, the absorption of MAPbI$_3$ in visible region has red shift.[38] In this work, we systematically study the changes of the optical absorption properties of MAPbI$_3$ under triaxial and biaxial strains.

MAPbI$_3$ adopts a tetragonal structure ($a = b \neq c$) at room temperature and its light absorption properties are different at equatorial plane and apical plane directions. The influences of triaxial and biaxial strains on the structure are also different. Therefore, the light absorptions in both equatorial plane and apical plane directions are considered under the external strains. The optical absorptions of MAPbI$_3$ under triaxial strain are shown in Fig. 8. The red shift of light absorptions at equatorial plane direction occurs at the absorption onset under compressive strains (< 4%, as shown in Fig. 8a), which is consistent with the experimental results.[38] It is also noteworthy that when -4% triaxial strain is applied, there are obvious peaks at 400 and 500 nm, which indicates the enhancement of the optical absorption capacity.[39] Under tensile strain, the blue shift of light absorption of MAPbI$_3$ at the absorption onset occurs. In the apical plane direction (Fig. 8b), the absorption spectra of MAPbI$_3$ at the absorption onset show red shift under compressive strain (< 4%) and blue shift under tensile strain, but blue shift occurs under 4% compressive strain. This is due to the destroying of the octahedron caused by the longer Pb-I bond under 4% strain (Fig. 3).

The changes of optical properties of MAPbI$_3$ under biaxial strain are shown in Fig. 9. In the equatorial plane (see Fig. 9a), the light absorption spectrum of MAPbI$_3$ show red shift at the absorption onset under compressive strain, and its light absorption capacity increases. The absorption spectrum of MAPbI$_3$ under biaxial tensile strain is blue-shifted at the absorption onset. In the apical plane (see Fig. 9b), the absorption spectra of MAPbI$_3$ at the absorption onset shift blue under compressive strain, and shift red under tensile strain (up to 2%).

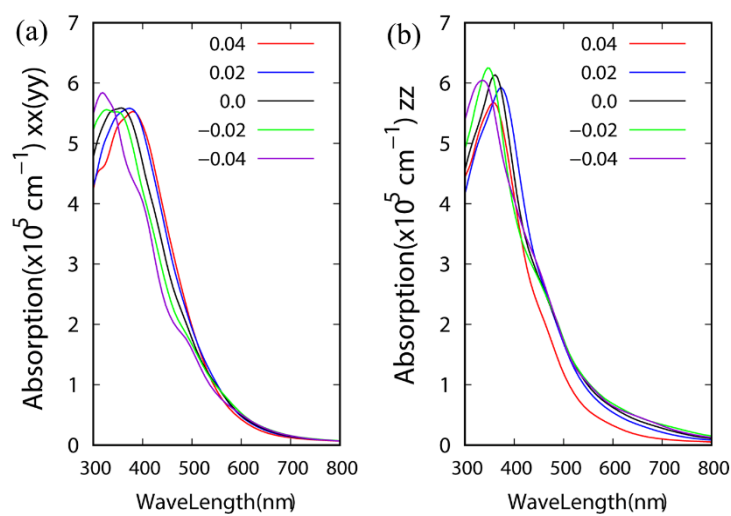

Figure 8. Optical absorption of MAPbI$_3$ (a) in plane and (b) out-plane under triaxial strain

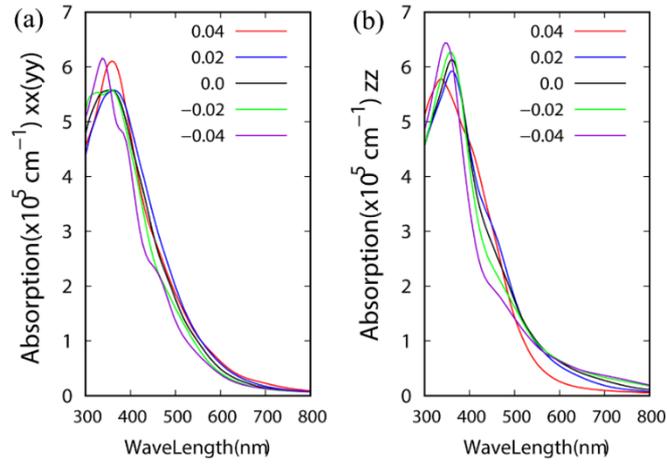

Figure 9. Optical absorption of MAPbI$_3$ (a) in plane and (b) out-plane under biaxial strain

## 6. Band structures and density of states under strain

MAPbI$_3$ possesses excellent optoelectronic properties, such as suitable direct bandgap, long carrier lifetime and diffusion length. The band gap of the material is closely related to its PCE. According to Shockley-Queisser limitation[40], if the bandgap of OIHPs can be tuned to 1.2~1.4 eV, the PCE will theoretically reach 33%. Therefore, reducing the bandgap of MAPbI$_3$ is an effective way to improve its PCE. Du *et al*. applied hydrostatic pressure to the cubic phase structure of MAPbI$_3$ by the first-principles calculations. It was found that the bandgap of MAPbI$_3$ could be reduced to 0.08 eV.[41] Here, we systematically study the effects of triaxial and biaxial strains on the electronic structure of tetragonal phase MAPbI$_3$.

The 1s orbital of the Pb atom was chosen as the benchmark to align the band for comparing the changes of the band structure under strain. The band gaps and band edge positions of MAPbI$_3$ under triaxial strain are shown in Fig. 10. We can see that tensile strain increases the energy of valence and conduction band, while compression strain decreases the energy of valence and conduction band. However, the effect of strain on the conduction band is more significant than that on the valence band, which increases the band gap under tensile strain but decreases it under compressive strain. The anomaly is that the variation of the valence band (-1.271 eV) is larger than that of the conduction band (-1.251 eV) at -4% compressive strain, which increases the bandgap of MAPbI$_3$ to 1.632 eV. Previous studies[42, 43] have shown that the shortening of the Pb-I bond in the crystal structure of MAPbI$_3$ enhances the orbital overlap and the band dispersion, which leads to the decrease of the band gap. The larger the tilt degree of [PbI$_6$] octahedron (i.e. the smaller bond angle of Pb-I-Pb) suggests the weaker the orbital overlap between Pb and I atoms and the smaller band dispersion, which leads to the increase of band gap. The relationship between the Pb-I bond (or the Pb-I-Pb bond angle) of MAPbI$_3$ and the bandgap under triaxial strain were shown in Fig. 11. We can see that the Pb-I bonds ($l_e$, $l_a$) and the Pb-I-Pb bond angles ($\alpha_e$, $\alpha_a$) increase under tensile strain (leading to larger band gap) but decrease under compressive strain (leading to smaller band gap), which indicates that the effect of Pb-I bond on the band gap of MAPbI$_3$ under triaxial strain is greater than that of Pb-I-Pb bond angle. However, the effect of Pb-I-Pb bond angle on the band gap becomes greater than that of Pb-I bond at -4% compressive strain, which increases the band gap of MAPbI$_3$.

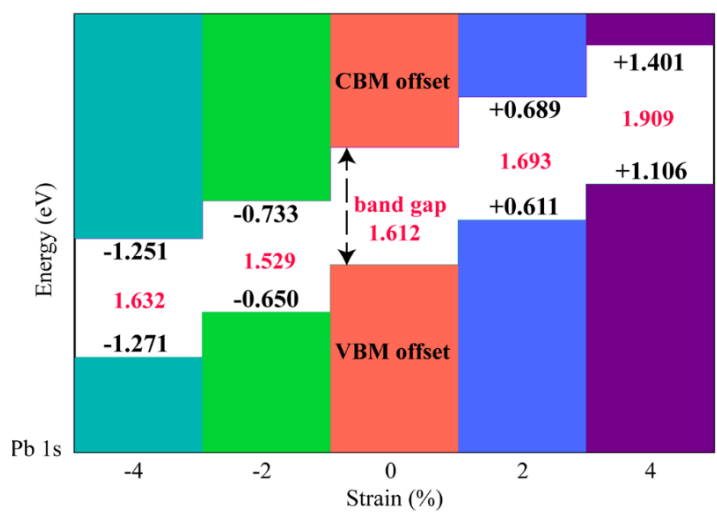

Figure 10. The changes of band gap and band edge position of MAPbI$_3$ under triaxial strain.

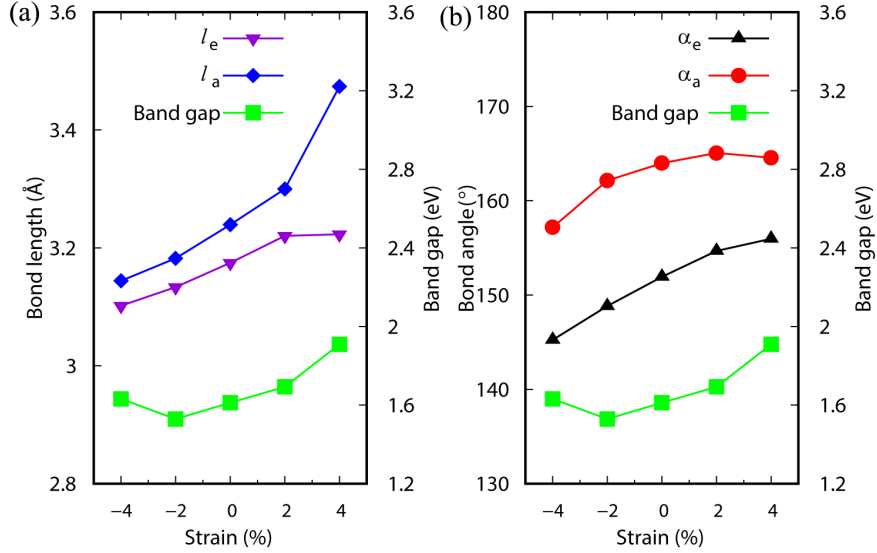

Figure 11. The relationship between Pb-I bond (or Pb-I-Pb bond angle)of MAPbI$_3$ and the band gap under triaxial strain. $l_e$, $l_a$ represent the Pb-I bond in the equatorial and apical plane, respectively. $α_e$, $α_a$ represent the Pb-I-Pb bond angles in the equatorial and apical plane, respectively.

The band gaps and edge positions of tetragonal MAPbI$_3$ under biaxial strain are shown in Fig. 12. The band gap of MAPbI$_3$ increases under biaxial tensile strain and decreases under compressive strain. The band gap of MAPbI$_3$ under 2% tensile strain decreases by 0.033 eV compared with that under 4% tensile strain, while it increases slightly (0.015 eV) at -4% compressive strain compared with that under -2% compressive strain (see the next section for this anomaly).

Under triaxial and biaxial strains, the changes of the Pb-I bond length and Pb-I-Pb bond angle in the lattice structure of MAPbI$_3$ are similar in equatorial plane but different in apical plane. Interestingly, their band gaps gradually increase from -2% to 2% triaxial or biaxial strain (see Fig. 11 and Fig. 13), indicating that the band gap of MAPbI$_3$ is mainly dominated by equatorial plane Pb-I bond in the small strain range. With the increase of compressive strain, the effect of Pb-I-Pb bond angle on the band gap of MAPbI$_3$ increases gradually, leading to the increase of the band gap under -4% triaxial and biaxial compressive strain. The band gap increases to 1.632 eV under -4% triaxial compressive strain because the Pb-I-Pb bond angle in the apical plane ($α_a$) also affects the band gap. Under 4% biaxial strain, the band gap of MAPbI$_3$ decreases by 0.033 eV compared with that under 2% biaxial strain, which may be due to the influences of Pb-I bonds in apical plane ($L_a$) on the band gap. For further verification, the electronic density of states of MAPbI$_3$ under strain will be calculated and discussed in the next section

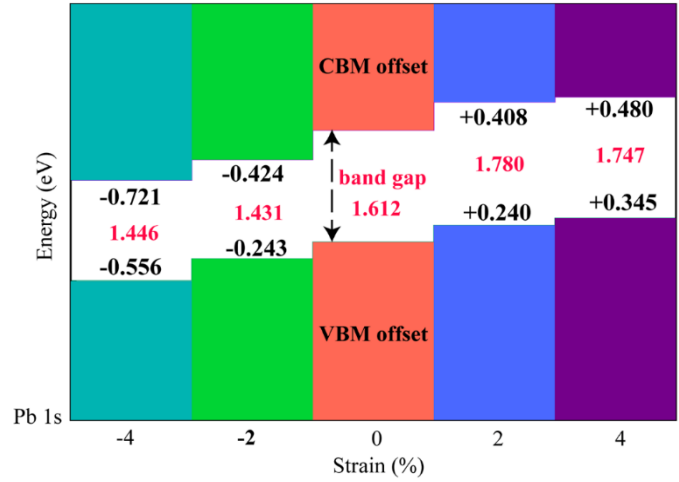

Figure 12. The changes of band gap and band edge position of MAPbI$_3$ under biaxial strain.

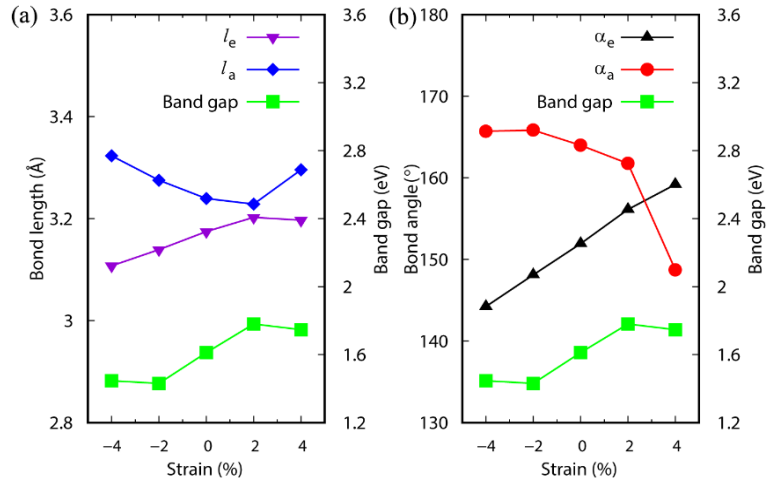

Figure 13. The relationship between Pb-I bond (or Pb-I-Pb bond angle) of MAPbI$_3$ and the band gap under biaxial strain. $l_e$, $l_a$ represent the Pb-I bond in the equatorial and apical plane, respectively. $α_e$, $α_a$ represent the Pb-I-Pb bond angles in the equatorial and apical plane, respectively.

The partial density of states of MAPbI$_3$ under triaxial and biaxial strains are shown in Fig. 14 and Fig. 15 (I$_e$ and I$_a$ represent I atoms in the equatorial and apical plane, respectively). It can be seen that the valence bands are mainly contributed by the *p* orbital of the I atoms and the *s* orbital of the Pb atoms, while the conduction band is mainly contributed by the *p* orbital of the Pb atoms and the *p* orbital of the I atoms, which is consistent with previous work.[44]

It is noteworthy that the valence band is mainly composed of the *s* orbital of Pb atom and the *p* orbital of the $I_e$ atom, leading to similar changes of band gap under triaxial and biaxial strains. Under 4% biaxial strain, the contribution of the *p* orbital of the $I_a$ to the valence band is equivalent to that of $I_e$ (Fig. 15), which indicates that the Pb-I bond in apical plane also affects the band edge positions of the $MAPbI_3$. This is why the bandgap of $MAPbI_3$ decreases by 0.033 eV under 4% biaxial tensile strain compared with 2% biaxial tensile strain.

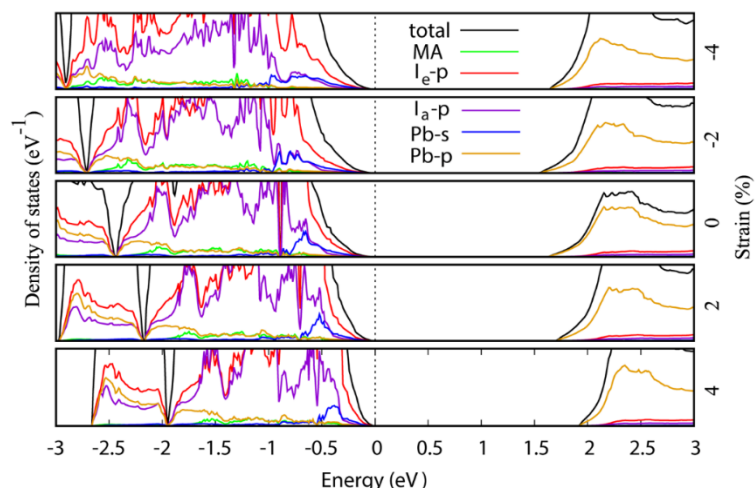

Figure 14. The partial density of states of $MAPbI_3$ under triaxial strain. $I_e$ represents I atom in the equatorial plane and $I_a$ represents I atom in the apical plane.

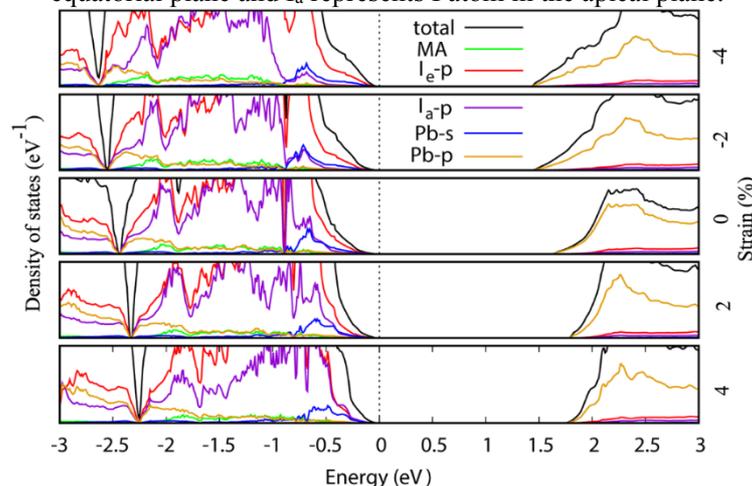

Figure 15. The partial density of states of $MAPbI_3$ under biaxial strain. $I_e$ represents I atom in the equatorial plane and $I_a$ represents I atom in the apical plane.

## 7. Conclusion

In summary, we investigated the structural stability and optoelectronic properties of tetragonal $MAPbI_3$ under external strain by using the first-principles calculations, and the mechanism was revealed by analyzing the changes of bond length and bond angle.

Our calculation results show that the migration barrier of I ions in the equatorial plane increases under compressive strain, which hinders the migration of I ions, enhancing the structural stability. Tensile strain decreases the migration barrier of I ions, which makes it easier for MAPbI$_3$ to decompose into MAI and PbI$_2$, reducing their lifetime. The light absorption of MAPbI$_3$ shows red shift in visible light under compressive strain, and the peak shows that the light absorption is enhanced under -4% compressive strain The band gaps of MAPbI$_3$ is mainly dominated by Pb-I bond in the equatorial plane in the small strain ranges, leading to gradual increase of the band gaps from -2% to 2% triaxial or biaxial strain. According to our calculations, it is suggested that when the -4% biaxial compressive strain is applied, not only increase the migration barriers of I ions, but also the band gap can be tuned to a suitable value, which is expected to further increase the power conversion efficiency of MAPbI$_3$. This will provide significant guidance for the commercial applications of MAPbI$_3$-based perovskite solar cells.

**AUTHOR INFORMATION**

**Corresponding Author**

*E-mail: hongjw@bit.edu.cn

**Acknowledgements**

This work is supported by the National Science Foundation of China (Grant No. 11572040), the National Materials Genome Project (2016YFB0700600) and the Thousand Young Talents Program of China. Theoretical calculations were performed using resources of the National Supercomputer Centre in Guangzhou which is supported by the Special Program for Applied Research on Super Computation of the NSFC-Guangdong Joint Fund (second phase) under Grant No U1501501.